\documentstyle[11pt]{article}
\renewcommand{\baselinestretch}{1.0}
\def\ds{\displaystyle}
\def\ra{{\bf r}_{\! \scriptscriptstyle A}}
\def\rb{{\bf r}_{\! \scriptscriptstyle B}}
\def\zrc{{\bf r}_{\! \scriptscriptstyle C}}
\def\rc0{{\bf r}_{{\! \scriptscriptstyle C}_0}}
\def\rcj{{\bf r}_{{\! \scriptscriptstyle C}_j}}
\def\tcj{\theta_{{\! \scriptscriptstyle C}_j}}
\def\dsp{d_{\scriptscriptstyle S}}
\def\rs{r_{\! \scriptscriptstyle S}}
\def\ep{{\mbox{\large e}}}
\newcommand{\bit}{\begin{itemize}}
\newcommand{\eit}{\end{itemize}}
\newcommand{\be}{\begin{equation}}
\newcommand{\ee}{\end{equation}}
\begin{document}
\begin{center}
{\huge Trace formula for an ensemble of bumpy billiards }

\vspace{1.0cm}

{\Large Nicolas Pavloff}
\end{center}

\vspace{0.5 cm}

\noindent Division de Physique Th\'eorique\footnote { Unit\'e de
Recherche des Universit\'es Paris XI et Paris VI associ\'ee au CNRS.},
Institut de Physique Nucl\'eaire, F-91406 Orsay Cedex, France \break

\vspace{0.5 cm}
\begin{center}
{\bf abstract}
\end{center}

	We study the semiclassical quantization of an ensemble of billiards
with a small random shape deformation. We derive a trace formula averaged over
shape disorder. The results are illustrated by the study of supershells in
rough metal clusters.

\vspace{5cm}
\noindent PACS numbers :\hfill\break
\noindent 03.65.Sq Semiclassical theories and applications.\hfill\break

\noindent IPNO/TH 94-69 \hspace{0.5cm} {\it submitted to J. Phys.
A: Math. Gen. }\hfill\break
\newpage

\section{Introduction}

	Quantum billiards have been extensively studied as model systems having
a chaotic or integrable classical dynamics (see {\it e.g.} \cite{Gu90}). They
have also been considered as simple models for atomic nuclei or metal clusters.
More recently they have been studied experimentally in ballistic
microstructures
\cite{Mar92,Lev93} and in microwave cavities \cite{micro}. In most of the
experimental studies the shape of the equivalent billiard is determined only on
the average. For instance, metal clusters have an underlying ionic
background which implies an unavoidable degree of roughness of order of the
interatomic distance, {\it i.e.} of order of the Fermi wavelength. For
microstructures the roughness is mainly due to the irregularities in the
depletion layer which can be estimated of being also of order of the Fermi
wavelength \cite{bouchiat}. Moreover clusters are produced in large amounts in
molecular beams and one has to consider an ensemble of shapes. In the same line
one also frequently considers an ensemble of microstructures (typically $\sim
10^5$) with a size dispersion ranging from $2 \%$ \cite{Reu95} up to $30 \%$
\cite{Lev93}.

	The mean free path in the experiments quoted above is larger than the
typical distances in the system, thus the billiard model is still meaningful.
Nevertheless it should be corrected due to shape irregularities. In this
paper we make an attempt of studying this phenomenon by considering an ensemble
of billiards (in any dimension) obtained by a random deviation from a fixed
initial shape (hereafter denoted as the perfect or unperturbed shape). We speak
below of rough or bumpy billiards.

	The paper is organized as follows : in Sec. 2 and 3 we derive a
semiclassical trace formula averaged over the ensemble of rough billiards. As
an illustration the method is applied in Sec. 4 to study the supershell
oscillations in rough metal clusters. We give our conclusions and compare with
previous works in Sec. 5.

\section{Green function in presence of shape disorder}

	Modern semi-classics have made extensive use of trace formulae such as
derived by Gutzwiller in the context of quantum chaos (see \cite{Gu90} and
references therein) or for quantum billiards by Balian and Bloch \cite{BB72}.
In this approaches the level density is obtained by computing the trace of the
Green function $G(\rb, \ra, k)$, solution of the Helmholtz
equation with Dirichlet or Neumann boundary conditions. It is written in the
semi-classical limit as a sum of contributions arising from the classical
orbits of the system. In the case of a billiard it reads schematically :

\be\label{e2}
G(\rb, \ra, k) \; = \sum_{A \rightarrow B}
{\cal D}(k) \ep^{\ds i(k L - \mu\pi/2)} \; .
\ee

\noindent where the sum is taken over the classical trajectories going from
point A to point B. In (\ref{e2}) $k$ is the wave-vector, it is related to the
energy by $\hbar k=\sqrt{2mE}$ (or $\hbar k=E/c$ for microwave cavities) ;
$\mu$ is a Maslov index and ${\cal D}(k)$ an amplitude characterizing the
trajectory of length $L$ considered. A general expression of ${\cal D}$ can be
found in \cite{Gu90,BB72}.

	Let us now treat the case of the rough billiard. We consider that
the shape disorder is weak enough so that a point $\zrc$ on the frontier of
the bumpy billiard can be written unambiguously as :

\be\label{e3} \zrc = \rc0 + h( \rc0) \;  {\bf n}_0 \; , \ee

\noindent where $\rc0$ is a point of the frontier of the perfect billiard and
${\bf n}_0$ the normal at this point. $h$ is a random displacement, the
characteristics of which will be specified later.

	Let us branch the perturbation (\ref{e3}) starting from a perfect
billiard. The direct orbit going from $A$ to $B$ without bouncing on the
boundary is not affected. If the shape modification is small enough (this will
be made mathematically precise later) orbits experiencing only a few bounces
will not be drastically altered (see Fig. 1). At first order in the
semi-classical approximation one will consider that only their change in length
is of importance, because it appears in the rapidly oscillating term $\exp (i k
L)$ of (\ref{e2}). The modification of the slowly varying amplitude ${\cal
D}(k)$ is simply neglected. Long orbits on the contrary experience many bounces
and they may be completely different in the rough enclosure and the perfect
billiard. They also will be drastically different from a bumpy billiard to
another, and ensemble averaging will very efficiently damp their contribution.
Hence it is legitimate in the semi-classical limit to work in a perturbative
approach where the only extra contribution with respect to the perfect billiard
is the modification of the length of the orbits in (\ref{e2}). The spirit of
the present approach is very common (see {\it e.g.} \cite{Zim62,JRU94,Cre95})
and the results are similar to those obtained by other techniques (see below).

	If we denote by $\delta L$ the modification of length of an orbit due
to surface roughness, using the notations of Fig. 1, for a single bounce
trajectory one obtains :

\be\label{e4}
\delta L = (A E + E B) - (A C_0+C_0 B) \simeq (A C + C B) - (A C_0 +C_0 B)
\; . \ee

	In (\ref{e4}) $C_0$ (resp. $E$) is the point of specular reflection in
the perfect (resp. bumpy) billiard. Since $A E+E B$ is an extremum
of the length, at first order it can be computed replacing $E$ by a nearby
point. This has been done in the {\it r.h.s.} of (\ref{e4}) where point
$C$ was used, $C$ being the intersection of the normal to the
perfect billiard at $C_0$ with the bumpy frontier (see Fig. 1). This can be
easily extended to orbits with $n$ bounces and simple geometry yields

\be\label{e5}
\delta L \simeq 2 \sum_{j=1}^{n} h(\rcj) \cos \tcj \; ,
\ee

\noindent where the sum is extended over all the bouncing points $C_j$ of the
classical trajectory on the boundary of the perfect billiard ($\tcj$ is the
normal angle of incidence at point $C_j$, see Fig. 1).

	Then the semi-classical Green function in the rough billiard is written
as

\be\label{e6}
G(\rb, \ra, k) \; \simeq \left.\sum_{A\rightarrow B}\right.^{\!\!\! 0}
{\cal D}(k) \ep^{\ds i(k L - \mu\pi/2)}
\exp \left( 2 i k \sum_{j=1}^{n} h(\rcj) \cos\tcj\right)
\; , \ee

\noindent where the upper index $0$ indicates that the sum is taken over the
trajectories of the {\it perfect billiard}.

	Careful derivation puts the following limitations to the use of Eqs.
(\ref{e5}) and (\ref{e6}) :
\bit
\item[(a)]  $|h|/L \ll |\nabla h| \ll 1$, or in other words : $L$ should be
greater than the typical distance between two bumps, itself being greater than
the amplitude of shape disorder. These restrictions insure that replacing $E$
by $C$ in (\ref{e4}) is legitimate and that $\delta L \ll L$.
\item[(b)] $k |h|^2 \ll L$ insures that using the approximate length $L +
\delta L$ in the semi-classical formula yields corrections which are indeed
small compared to the main term (\ref{e6}).
\item[(c)] One should also make sure that diffractive corrections to the
leading order semi-classics can be safely neglected. Hence, the typical
distance between two bumps ($|h|/|\nabla h|$) should be larger than the
wave-length ($1/k$). If not, the amplitude ${\cal D}(k)$ is significantly
modified by the surface roughness.
\eit

	Keeping in mind the physical examples given in the introduction one
sees that among the above restrictions only the ones involving $\nabla h$ are
not trivially satisfied. Indeed in the case of a large shape disorder
the distance between two bumps is of order of the amplitude of a
bump (then $\nabla h$ is of order 1) and also diffractive effects will have to
be taken into account. Hence (\ref{e6}) is rigorously applicable only for small
roughness (characterized by the restrictions (a), (b) and (c)). Nevertheless we
will see in section 3 that in this limit the effects of the surface roughness
are already very noticeable. Hence one can argue that when (\ref{e6}) is no
longer valid the associated oscillations in the level density are already
almost completely damped (see (\ref{e12})).

	We recall that (\ref{e6}) is only valid for short orbits. The
contribution of long orbits in the bumpy billiard cannot be inferred from the
motion in the perfect billiard. In order to have a formula rigorously
applicable let us now damp the contribution of long orbits by performing an
ensemble average of the Green function. The computation is straightforward and
the average quantity reads

\be\label{e7}
< G(\rb \ra, k) > \; \simeq  \left.\sum_{A\rightarrow B}\right.^{\!\!\! 0}
{\cal D}(k) \ep^{\ds i(k L - \mu\pi/2)} \prod_{j=1}^{n} \chi (2 k \cos\tcj)
\; , \ee

\noindent where $\chi$ is the characteristic function of the rough shape
\cite{Ogi91}. It is the Fourier transform of the probability density of the
frontier displacement $h$ :

\be\label{e7x} \chi (s) = \int_{-\infty}^{+\infty} p(h)\ep^{\ds i s h} d h
\; .\ee

	In (\ref{e7}) we have made the hypothesis that the bounces were
separated by a distance larger than the correlation length of the shape
disorder, {\it i.e.} $p(h({\bf r}_{{\! \scriptscriptstyle C}_1}), ... ,
h({\bf r}_{{\! \scriptscriptstyle C}_n}))=
p(h({\bf r}_{{\! \scriptscriptstyle C}_1}))\times ... \times
p(h({\bf r}_{{\! \scriptscriptstyle C}_n}))$. This restriction is not necessary
but it simplifies the presentation. Following Ref. \cite{Cre95} one could think
of deformations strongly violating this assumption : this would be the case for
instance of an unperturbed circle going to a rough ellipse. In this case it
will
always be possible to use the present formalism if considering the perfect
ellipse as the unperturbed billiard.

	Most of the authors (see \cite{Ogi91}) choose a gaussian distribution
for $h$ with standard deviation $\sigma$ :

\be \label{e8}
p(h) = {\ds 1\over \ds \sigma \sqrt{2\pi}}
\exp ( -{\ds h^2\over \ds 2\sigma^2} ) \qquad \hbox{and} \qquad
\chi(s) = \exp ( -{\ds s^2\sigma^2 \over \ds 2} ) \; . \ee

	Note here that on the basis of Eq. (\ref{e8}) only one cannot check the
validity of the restrictions (a), (b) and (c) above : they mainly concern the
correlation length of the random function $h$ and not only the characteristic
function $\chi$ which is our unique ingredient. The restrictions will be
fulfilled if $\sigma$ is smaller than the correlation length, itself being
smaller than typical distances in the billiard.

	The physics embodied in Eq. (\ref{e7}) can be simply interpreted by
noticing that the wave pro\-pa\-gates as in a perfect billiard with, at each
bounce, an extra damping factor $\chi (2k\cos \tcj)$ ({\it i.e.}
$\exp(-2 k^2 \sigma^2 \cos^2 \tcj)$ in the gaussian model (\ref{e8})). The
general form $\chi (2k\cos \tcj ) $ is commonly obtained in Kirchhoff theory of
wave scattering from rough surfaces \cite{Ogi91}. As anticipated this damps
very efficiently the contribution of orbits experiencing many
reflections. The quantity $k \sigma \cos \tcj$ is known as the Rayleigh
parameter \cite{Ray45} and characterizes to which extend an incident wave is
sensitive to surface roughness : as one would intuitively expect, the
sensitivity is maximum for perpendicular incidence ($\cos \tcj \simeq 1$) and
short wave lengths ($k\sigma \gg 1$).

\section{Trace formula in the bumpy billiard}

	The next step in the derivation of a trace formula is to compute
the level density by taking the trace of the Green function :

\be\label{e9}
\rho (k) = -{2\dsp k\over \pi} \; \hbox{Im} \; \int d^Dr \;
G({\bf r}, \; {\bf r}, k) =
= -{2\dsp k\over \pi} \; \hbox{Im} \; \hbox{Tr} \; \widehat{G}(k)
\; . \ee

	In (\ref{e9}) the integral extends over the interior of the billiard
and $D$ is the dimension of space. $\widehat{G}(k)$ is the operator whose
matrix elements in configuration space give $G(\rb, \ra, k)$. $\dsp$ accounts
for a possible spin degeneracy (in this case $\dsp =2$, $\dsp=1$ otherwise).

	It is customary to separate the contribution of the quasi zero length
orbits to which the semi-classical approximation (\ref{e2}) does not apply.
These orbits contribute to the smooth part $\bar\rho (k)$ of the level density
through the ``Weyl expansion" (see {\it e.g.} \cite{Bal76}). In three
dimensions and for Dirichlet boundary conditions it reads

\be\label{e10} \bar\rho (k) = \dsp \left( {V k^2\over 2 \pi^2} -
{S k\over 8\pi} + ... \right) \; , \ee

\noindent where $V$ is the volume of the billiard and $S$ its surface area. In
the typical case of a bumpy sphere of radius $R$ with disorder of type
(\ref{e8}) the average $<\bar\rho (k)>$ is easily computed ; one obtains
$<V>=4\pi (R^3/3+\sigma^2 R)$ and $<S>=4\pi (R^2+\sigma^2)$. $\sigma$ is
supposed to be small compared to $R$, thus surface roughness poorly affects the
smooth part of the spectrum : for practical computations we will approximate
$<\bar\rho(k)>$ by the value $\bar\rho(k)$ in the perfect billiard. We have
given here a generic 3 dimensional example but the same holds in any
dimension.

	A quantity of primary interest is the oscillatory part
$\widetilde\rho(k)$ of the level density. As shown by Gutzwiller \cite{Gu90}
and Balian and Bloch \cite{BB72}, inserting the semi-classical Green function
(\ref{e2}) in Eq. (\ref{e9}) and performing a non trivial stationary phase
analysis leads to a trace formula which reads schematically :

\be\label{e11} \widetilde\rho (k) \; = \sum_{PO} {\cal A}(k)
\sin(k L+\nu\pi /2) \; . \ee

	The sum (\ref{e11}) extends over all the classical periodic orbits
(PO's) of the system. As in (\ref{e2}) ${\cal A}(k)$ is a slowly varying
amplitude and $\nu$ a Malsov index characteristic of the PO of length $L$
considered (see \cite{Gu90,BB72}). Some PO's may form continuous families, {\it
i.e.} some orbits -- forming a continuous set -- may all have the same length
and the same topology (such as the bouncing ball orbit in the stadium billiard
or the PO's in integrable enclosures). Two orbits of the same family differ
only by their bouncing points. Each family is represented by a single term in
the summation (\ref{e11}) but its amplitude is enhanced by additional powers of
$k$ with respect to the contribution of an isolated orbit. This is related in
\cite{Cre91} to local (possibly global) continuous symmetries of the
hamiltonian.

	If one wishes now to write a trace formula for a rough billiard,
contrary to what happens for the Green function, the form (\ref{e11}) of
$\widetilde\rho (k)$ is different in the bumpy and in the perfect billiard. The
reason is that some PO's may appear in continuous families and roughening
destroys these families. This can be illustrated on the following two
dimensional example : consider a rectangular billiard of which one edge --the
upper one say-- has been modified to adopt a sinusoidal shape. One of the
important continuous families of PO's in the perfect rectangle --the vertical
bouncing ball-- is reduced in the bumpy rectangle to only a couple of orbits
(those hitting the sinusoidal upper edge at points with horizontal tangent).
Nevertheless the level densities of the two systems are certainly very similar
if the edge deformation is small. This type of problem has been recently
addressed in Ref. \cite{Cre95,Pri94} and deserves a careful treatment. For
isolated orbits it might in some cases be explained semiclassicaly by the
introduction of complex PO's in the trace formula (see the discussion in
\cite{BB72} and also \cite{PA94}). However we can bypass this kind of
subtleties when averaging over disorder because it is legitimate to permute the
trace and the average : $ < \hbox{Im} \; \hbox{Tr} \; \widehat{G}(k) > =
\hbox{Im} \; \hbox{Tr} < \widehat{G}(k) >$. Hence $<\widetilde\rho (k) >$ can
simply be computed by inserting the average $<G({\bf r}, \; {\bf r}, k)>$ in
the trace (\ref{e9}). Since $<G({\bf r}, \; {\bf r}, k)>$ is evaluated by using
the orbits of the perfect billiard, the saddle point can be performed in the
usual manner and yields the average oscillating part of the level density :

\be\label{e12}
<\widetilde\rho (k) > \; \simeq \left.\sum_{PO}\right.^{\! 0}
{\cal A}(k) \sin(k L+\nu\pi /2) \prod_{j=1}^{n} \chi (2 k \cos\tcj)
\; . \ee

	The index $0$ in the summation indicates as before that all the
quantities are evaluated in the unperturbed billiard. Hence $n$ in (\ref{e12})
is the number of bounces of a PO in the perfect billiard, the $\theta$'s are
the normal angle of incidence.

	Formula (\ref{e12}) is the most important result of the paper. It is
valid for rough billiards in any dimension. It is interesting to note
that when considering an integrable perfect billiard with {\it an ergodic
perturbation}, the contribution of short orbits is correctly accounted for by
Eq. (\ref{e12}), even without ensemble averaging. By ergodic, we mean that
``any statistical average taken over many different parts of one shape
realization is the same as an average over many realizations'' \cite{Ogi91}. In
the case of integrable unperturbed billiards, all the orbits occur in families
and the spatial integration (\ref{e9}) over a continuous family is -- by the
hypothesis of ergodicity -- equivalent to ensemble averaging (this was
certainly not the case in the simplified example above of sinusoidal
deformation of a rectangle billiard because sine is not an ergodic function).
We recall that this is not correct for long orbits which may be very different
in the bumpy and the perfect billiard, in this case ensemble average is
necessary to damp the associated oscillations.

	To fix the ideas we apply (\ref{e12}) to a bumpy sphere of radius $R$.
The total average level density $<\!\rho\! > \simeq \bar\rho + <\widetilde\rho
>$ is plotted on Fig. 2 for two values of the surface roughness in the gaussian
model (\ref{e8}) : $\sigma = 0.04 \; R$ and $0.06 \; R$. The amplitude ${\cal
A}(k)$ for each PO in the perfect sphere can be found in \cite{BB72}. 75 PO's
were included, with a maximum length of 26 times the radius. More precisely in
the terminology of \cite{BB72} the maximum values of the parameters are $t=5$
and $p=15$ ($t$ being the winding number of an orbit around the center and $p$
the number of bounces). We also indicate with black arrows the location
of the first 15 eigenlevels in the unperturbed sphere. One sees on the figure
that in the lower part of the spectrum, the wavelength being large, surface
irregularities do not perturb the eigenstates much and there is still a strong
bunching of levels. This shell effect gradually disappears for increasing
energies (when the wavelength becomes comparable with the amplitude of the
disorder).

\section{Supershells in rough metal clusters}

	In this section we illustrate the above results with the example of
shell structure in metal clusters. Rough clusters will be described by a simple
model first introduced in Ref. \cite{Rat80} : $N$ electrons are moving
independently in a bumpy sphere. The radius $R$ of the perfect sphere scales
with $N$ so that the mean electronic density is kept constant and equal to its
bulk value : $R = \rs N^{1/3}$, $\rs$ being the Wigner-Seitz radius of the bulk
material. The standard deviation from this average shape is of order of atomic
distances {\it i.e.} of order $\rs$. Note that such irregularities are to be
taken into account even if the cluster is ``liquid-like" : the mean velocity of
the ionic cores is always by several orders of magnitude smaller than the
typical electronic Fermi velocity. Hence, as far as electronic motion is
concerned, the ionic cores can be considered as frozen and this automatically
implies a certain degree of surface roughness.

	In the unperturbed sphere, due to the high symmetry of the potential
there is a strong bunching of levels leading to shell structure and magic
numbers (see {\it e.g.} \cite{BM75}). This shell structure is modulated when
the cluster size grows (it first disappears, then increases, {\it etc}...).
This is denoted ``supershell structure'' and was first noticed in Ref.
\cite{BB72} and seen experimentally by the Copenhagen group \cite{Ped91}. We
will here study this effect in a rough cluster.

	One of the most important observables when studying shell structure is
the shell energy which is the oscillating part of the total energy. Shells are
experimentally detected on a mass spectrum ; roughly speaking clusters with
relatively smaller total energy (corresponding to minima of the shell energy)
are more stable and will be more numerous in a beam. This is also correlated
with larger ionization potential, but for this observable shell effects
decrease with cluster size, making its study more difficult.

	The Fermi wave vector $k_f$ and total energy $E_{tot}$ are defined by :

\be\label{e13} N = {\cal N}(k_f) = \int_0^{k_f} \rho(k) d k \qquad \hbox{and}
\qquad E_{tot}(N) = \int_0^{k_f} {\ds \hbar^2 k^2\over \ds 2m}
\rho(k) d k \; , \ee

	In (\ref{e13}) ${\cal N}(k)$ is the integrated level density, or
spectral staircase. As the level density, ${\cal N}$ can be written as the sum
of a smooth quantity $\bar{\cal N}$ and an oscillating part $\widetilde{\cal
N}$. The same holds for $k_f$ and $E_{tot}$ considered as functions of $N$. The
smooth terms are defined by :

\be\label{e15}  N = \bar{\cal N}(\bar{k}_f) = \int_0^{{\bar k}_f} \bar{\rho}
(k) dk \qquad \hbox{and} \qquad
\bar{E}_{tot}(N) = \int_0^{{\bar k}_f} {\hbar^2 k^2\over 2m}
\bar{\rho} (k) dk \; . \ee

	As explained before we will identify the average smooth quantities
$<\bar{E}_{tot}>$ and $<\bar{k}_f>$ with their value in the perfect sphere. On
the basis of the Weyl expansion in the sphere and of (\ref{e15}) one obtains
the following relations :

\be\label{e15b} \bar{k}_f(N) = \kappa_f \left[ 1 - {\ds a_2\over\ds 3 a_3}
\left( {N\over \ds a_3} \right)^{-1/3} + {\ds a_2^2-3a_1a_3\over\ds 9a_3^2}
\left( {N\over \ds a_3} \right)^{-2/3} + \cdots \right] \; , \ee

\noindent and

\be\label{e16} \bar{E}_{tot} (N) = \varepsilon_f \left[ {\ds 3 a_3\over \ds 5}
\left( {N\over \ds a_3} \right)       - {\ds a_2\over \ds 2}
\left( {N\over \ds a_3} \right)^{2/3} + {\ds a_2^2-2a_1 a_3\over \ds 3 a_3}
\left( {N\over \ds a_3} \right)^{1/3} + \cdots \right] \; , \ee

\noindent where $a_3=2\dsp/9\pi$, $a2=-\dsp/4$, $a_1=2\dsp/3\pi$ (see
\cite{Bal76}). $\kappa_f$ and $\varepsilon_f$ are the bulk Fermi wave-vector
and Fermi energy : $\kappa_f\rs = a_3^{-1/3}$ and $\varepsilon_f = \hbar^2
\kappa_f ^2 /2m$.

	The average oscillating part is obtained by subtracting (\ref{e16}) to
the average total term. It is computed using the following approximation :

\be\label{e18} <\widetilde{E}_{tot}> =
<\int_0^{k_f} {\ds \hbar^2 k^2\over \ds 2m}\rho(k) d k> - \bar{E}_{tot}
\simeq \int_0^{<k_f>} {\ds \hbar^2 k^2\over \ds 2m}
<\rho(k)> d k - \bar{E}_{tot}\; . \ee

	The r.h.s. of (\ref{e18}) can be considered as a simple first
approximation of the exact result. It is made necessary by the difficulty
mentioned above for estimating $\rho(k)$ in an individual rough cluster. We can
invoke the hypothesis of ergodicity of surface disorder to make this
approximation sound : for a cluster such as created in a molecular beam one can
argue that the individual level density will have a pattern very similar to the
one displayed in Fig. 2 ({\it cf} the discussion in Sec. 3).  Hence, although
in a given cluster $\rho(k)$ is exactly a sum of delta pics, the bunching of
levels in an individual spectrum will disappear at the same wavelength that
predicted in the average $<\rho(k)>$.

	Then $<\!\widetilde{E}_{tot}\! >$ can be computed with the following
scheme : one first determines $<\! k_f\! >$ by numerical inversion of the first
term of Eq. (\ref{e13}) (with $<\!\rho\! >$ replacing $\rho$) and the integral
of the second term is then computed numerically. One can also evaluate the
asymptotic behaviour of the integrals (\ref{e13}) to get an analytical
expression. For this purpose the exact Fermi wave-vector in written as
$k_f=\bar{k}_f+\widetilde{k}_f$, where $\bar{k}_f$ is given by (\ref{e15b}) and
$\widetilde{k}_f$ is supposed to be small. Then performing on ${\cal
N}=\bar{\cal N}+\widetilde{\cal N}$ a first order limited expansion one gets

\be\label{b3} {\cal N}(k_f) \simeq \bar{\cal N}(\bar{k}_f) +
\widetilde{k}_f \left( {\ds d\bar{\cal N}\over d k}\right)_{\bar{k}_f}
+ \widetilde{\cal N}(\bar{k}_f) +
\widetilde{k}_f \left( {\ds d\widetilde{\cal N}\over d k}\right)_{\bar{k}_f}
\; . \ee

	Since ${\cal N}(k_f)=\bar{\cal N}(\bar{k}_f)=N$ by definition (see
(\ref{e13},\ref{e15})) one obtains :

\be\label{b4}
\widetilde{k}_f \simeq - \; {\ds \widetilde{\cal N}(\bar{k}_f)\over
\ds\bar{\rho}(\bar{k}_f)+\widetilde{\rho}(\bar{k}_f)}
\qquad . \ee

	This expression can be evaluated by keeping only the leading order of
$\widetilde{\cal N}$ (obtained by integration of (\ref{e11})) and
neglecting the oscillatory term at the denominator. Then averaging over
disorder yields :

\be\label{b5}
< \widetilde{k}_f > \simeq
{\ds 1\over \ds\bar{\rho}(\bar{k}_f)} \left.\sum_{PO}\right.^{\! 0}
{\ds {\cal A}(\bar{k}_f)\over\ds L} \cos(\bar{k}_f L + \nu\pi/2)
\prod_{j=1}^{n} \chi (2 \bar{k}_f \cos\tcj)
\; . \ee

	Then the shell energy $\widetilde{E}_{tot}$ is computed from the
difference between the total energy and its smooth part :

\begin{eqnarray}\label{b7}
\widetilde{E}_{tot} = E_{tot} - \bar{E}_{tot} & = &
\int_{\bar{k}_f}^{k_f} {\hbar^2 k^2\over 2m} \bar{\rho} (k) dk  +
\int_0^{k_f}           {\hbar^2 k^2\over 2m} \widetilde{\rho} (k) dk
\nonumber \\
 & \simeq & \widetilde{k}_f {\hbar^2 \bar{k}_f^2\over 2m}
\bar{\rho}(\bar{k}_f) + \int_0^{\bar{k}_f}
{\hbar^2 k^2\over 2m} \widetilde{\rho} (k) dk +
\widetilde{k}_f {\hbar^2 \bar{k}_f^2\over 2m} \widetilde{\rho}(\bar{k}_f) \; .
\end{eqnarray}

	In (\ref{b7}) we replaced the integrals by their large $N$
approximations. The dominant contribution is obtained by integrating by parts
the integral appearing in the last term of (\ref{b7}) ; it cancels due to
relation (\ref{b4}). After averaging over disorder the next order reads :

\be\label{b8}
<\widetilde{E}_{tot}> \simeq - {\ds \hbar^2\bar{k}_f\over\ds m}
<\widetilde{\cal F}(\bar{k}_f)> \simeq {\ds \hbar^2\bar{k}^2_f\over\ds 2 m}
\left.\sum_{PO}\right.^{\! 0}
{\ds 2 {\cal A}(\bar{k}_f)\over\ds \bar{k}_f L^2} \sin(\bar{k}_f L + \nu\pi/2)
\prod_{j=1}^{n} \chi (2 \bar{k}_f \cos\tcj) \; .\ee

	${\cal F}(k)$ is a primitive of ${\cal N}(k)$ which has been estimated
at first order in the {\it r.h.s.} of Eq. (\ref{b8}).

	This formula gives an accurate approximation of the value of
$<\widetilde{E}_{tot}>$ computed numerically : the result is shown on Fig. 3
for the gaussian model (\ref{e8}) and for several values of the surface
roughness  ($\sigma/\rs = 0$, 0.1, 0.2 and 0.3). The dashed line corresponds to
Eq. (\ref{b8}) and the solid line to numerical evaluation of the integrals
(\ref{e13},\ref{e18}). Here the integration adds an extra smoothing compared to
Fig. 2 and one needs to take into account much less orbits, the figure includes
35 PO's up to a length $L=12 R$. Actually a very reasonable result can be
obtained with only the 7 shortest orbits, we included more orbits here to have
an accurate description of shell effect in the supershell region $N^{1/3}\simeq
8$.

	We see on Fig. 3 that shell structure is very sensitive to surface
irregularities of small amplitude. Nevertheless roughness reduces all
oscillations without modifying the qualitative features of the supershells.
Hence the present approach legitimates the usual explanation of supershell
effects in rough metal clusters as being due to the interference of the square
and triangular orbits \cite{BB72,Ped91}, although these orbits might not exist
in an individual cluster. Including temperature effects and quantitatively
comparing with the experimental results could fix an order of magnitude for the
irregularities of the surface of large metal clusters. A very rough estimate
based on separation energies for small clusters gives the value $\sigma \sim
0.2 \; \rs$ \cite{moi}.

\section{Discussion}

	In this paper we have studied the oscillating part of the level density
in billiards with small size shape irregularities and we derived a
semiclassical trace formula averaged over shape disorder. The important feature
of the level density is the gradual disappearance of the oscillations with
increasing energy : when the wavelength is of order of the typical size of the
surface defects this induced a shift of the eigen-energies which leads after
averaging to a structureless level density.

	The same type of approach has recently been used to study, within
semiclassical approximation, the role of families of orbits broken by a small
perturbation of an initial shape having a local continuous symmetry. The
authors of Ref. \cite{Pri94} study the bouncing ball orbit in a deformed
stadium billiard and in Ref. \cite{Cre95} a general trace formula is derived
accounting for the role of broken families. In the present work, the main
difference is the inclusion of an ensemble average yielding a formula valid
also for isolated orbits. Moreover averaging damps the contribution of long
orbits to which the simple perturbation technique (\ref{e5},\ref{e6}) does not
apply. This averaging method is motivated by the experimental techniques of
mesoscopic and cluster physics.

	The method has been applied to study supershell oscillations in metal
clusters using a model accounting for the irregularities of the surface due to
the underlying ionic structure. It is also of interest in ballistics
microstructures with shape irregularities \cite{Bra95}.

	The problem of surface roughness of metal clusters has recently been
addressed using an other approach : in Ref. \cite{Aku94} the effect of disorder
is represented via addition to the hamiltonian of a random matrix perturbation
({\it cf} Ref. \cite{Aku93}). The results for the average level density and
shell oscillations are qualitatively very similar to what is presented here. In
addition the authors of Ref. \cite{Aku94} argue that these effects could
explain experimental shifts in the measured mass distribution. Note that in
Ref. \cite{Aku94} and also in the present study the effects of temperature are
indirect : although the usual temperatures reached in experiments are small
compared to the Fermi energy (one remains in the very degenerate limit
$k_{\! \scriptscriptstyle B} T \ll \varepsilon_f$) they are sufficient to
induce a disorder of the ionic arrangement which has a sizeable effect on shell
structure.

\

{\bf Acknowledgements:} It is a pleasure to thank V. M. Akulin, H. Bouchiat, M.
Brack, S. Creagh, P. Leb\oe uf, S. Reiman, K. Richter and C. Schmit for
fruitful discussions. I wish also to express my gratitude to D. Ullmo for
enlightening comments on semiclassical methods.

\newpage
\renewcommand{\baselinestretch}{.7}

\newpage

\vspace{1cm}
{\center {\Large {\bf Figure captions }}}
\vspace{1cm}

{\bf Figure 1.} ~Classical orbits going from point $A$ to point $B$ with one
bounce one the boundary. The thick lines correspond to the perfect billiard and
the thin lines to the bumpy billiard. Point $C_0$ (resp. point $E$) is the
point of specular reflection on the perfect (resp. bumpy) enclosure.
$\theta_{{\! \scriptscriptstyle C}_0}$ is the normal reflection angle at $C_0$.
$C$ is the intersection of the normal at $C_0$ with the bumpy boundary.

\vspace{1cm}

{\bf Figure 2.} ~Total average level density in bumpy sphere
for $\sigma = 0.04 \; R$ (solid line) and $\sigma = 0.06 \; R$
(thick solid line). The dashed line represents the smooth term $\bar\rho (k)$.
The black arrows indicate the location of the first 15 eigenlevels in the
perfect sphere.

\vspace{1cm}

{\bf Figure 3.} ~Average oscillating part of the total energy is rough metal
clusters as a function of $N^{1/3}$ for several values of surface roughness :
$\sigma/\rs = 0$ (upper graph), 0.1, 0.2 and 0.3 (lower graph).
$<\widetilde{E}_{tot}>$ is expressed in units of the bulk Fermi energy
$\varepsilon_f$. The solid line corresponds to numerical evaluation of the
integrals (\ref{e13},\ref{e18}) and the dashed line to Eq. (\ref{b8}).

\end{document}